\documentstyle[amssymb,prbbib,eqsecnum,aps]{revtex}
\newtheorem{theorem}{Theorem}[section]

\newtheorem{corollary}[theorem]{Corollary}

\newtheorem{definition}[theorem]{Definition}

\newtheorem{lemma}[theorem]{Lemma}

\newtheorem{proposition}[theorem]{Proposition}
\newtheorem{remark}[theorem]{\it Remark \rm }

\input cyracc.def
\font\eightcyr=wncyr10 scaled 800
\def\cyr{\eightcyr\cyracc}
\draft

\begin{document}
\title{Off-diagonal terms in symmetric operators \thanks{%
1991 Mathematics Subject Classification: 47A05, 47A66, 47B15. } \thanks{%
Key words and phrases: symmetric operators, Hilbert space, domain,
deficiency vectors, intertwining operators, operator matrices, raising and
lowering operators, second quantization.}}
\author{Palle E. T. Jorgensen \thanks{
E-mail: {\tt jorgen@math.uiowa.edu} } \thanks{%
Work supported in part by the NSF \#DMS-9700130. }}
\address{Department of Mathematics, The University of Iowa, Iowa
City, IA 52242-1419 U.S.A.}
\date{\today}
\maketitle

\begin{abstract}
In this paper we provide a quantitative comparison of two obstructions for a
given symmetric operator $S$ with dense domain in Hilbert space ${\cal H}$
to be selfadjoint. The first one is the pair of deficiency spaces of von
Neumann, and the second one is of more recent vintage: Let $P$ be a
projection in ${\cal H}$. We say that it is {\em smooth} relative to $S$ if
its range is contained in the domain of $S$. We say that smooth projections $%
\left\{ P_{i}\right\} _{i=1}^{\infty }$ {\em diagonalize} $S$ if\medskip 
\newline
(a)\quad $\left( I-P_{i}\right) SP_{i}=0$ for all $i$, and\newline
(b)\quad $\sup_{i}P_{i}=I$.\medskip \newline
If such projections exist, then $S$ has a selfadjoint closure (i.e., $\bar{S}
$ has a spectral resolution), and so our second obstruction to
selfadjointness is defined from smooth projections $P_{i}$ with $\left(
I-P_{i}\right) SP_{i}\neq 0$. We prove results both in the case of a single
operator $S$ and a system of operators.
\end{abstract}

\section{Introduction}

\label{Int}The following infinite-by-infinite matrices, also called infinite
tridiagonal matrices, 
\begin{equation}
\left( 
\begin{array}{cccccccc}
a_{1} & \bar{b}_{1} & 0 & 0 & \cdots & 0 & 0 & \cdots \\ 
b_{1} & a_{2} & \bar{b}_{2} & 0 & \cdots & 0 & 0 & \cdots \\ 
0 & b_{2} & a_{3} & \bar{b}_{3} &  & \vdots & \vdots &  \\ 
0 & 0 & b_{3} & a_{4} & \ddots &  & \vdots &  \\ 
\vdots & \vdots &  & \ddots & \ddots & \ddots &  &  \\ 
0 & 0 & \cdots &  & \ddots & a_{n} & \bar{b}_{n} &  \\ 
0 & 0 & \cdots & \cdots &  & b_{n} & a_{n+1} & \ddots \\ 
\vdots & \vdots &  &  &  &  & \ddots & \ddots
\end{array}
\right)  \label{eqInt.1}
\end{equation}
($b_{i}\in {\Bbb C}$, $a_{i}\in {\Bbb R}$) arise in the theory of moments, 
\cite{Akh65,Ham44} in noncommutative geometry,\cite{Con94,Arv99,Seg51,Far79}
and in mathematical physics.\cite{MaOr99,Sch99,Jor75,Jor76,Jor77} Such
matrices clearly define symmetric operators $S$ in the Hilbert space ${\cal H%
}=\ell ^{2}$, and it can be checked that the corresponding deficiency
indices (see (\ref{eqInt.7}) below) must be $\left( 0,0\right) $ or $\left(
1,1\right) $. These cases correspond to classical {\em limit-point,}
respectively {\em limit-circle,} configurations for the corresponding
generalized resolvent operators (see Refs.\ \CITE{Sto32,AkGl93}). The
limit-point case yields a selfadjoint closure $\bar{S}$, and we say that $S$
is {\em essentially selfadjoint.} This means that it has a spectral
resolution which is given by the spectral theorem applied to $\bar{S}$. In
the other case, there are nonzero vectors $x_{\pm }$ in $\ell ^{2}$ such
that 
\begin{equation}
\left\langle x_{\pm },Sy\pm iy\right\rangle _{\ell ^{2}}=0\text{\qquad for
all finite sequences }y.  \label{eqInt.2}
\end{equation}
It is also known\cite{Akh65} that $\bar{S}$ is selfadjoint if and only if $%
\sum_{n}\left| b_{n}\right| ^{-1}=\infty $, assuming that the numbers $b_{n}$
are all nonzero. (If some are zero, there is a natural modified condition.)

The most basic example of such infinite tridiagonal matrices from quantum
mechanics included the variables $p$, $q$ from Heisenberg's $pq-qp=\frac{1}{%
\sqrt{-1}}I$. To see this, realize $\ell ^{2}$ as $L^{2}\left( {\Bbb R}%
\right) $ via the orthonormal basis in $L^{2}\left( {\Bbb R}\right) $
consisting of the Hermite functions $h_{n}\left( \,\cdot \,\right) $. Then $%
pf\left( x\right) =\frac{1}{\sqrt{-1}}f^{\prime }\left( x\right) $ and $%
qf\left( x\right) =xf\left( x\right) $, say for $f$ in the Schwartz space $%
{\cal S}\left( {\Bbb R}\right) $, and the corresponding raising and lowering
operators are $\left( p+iq\right) h_{n}=\sqrt{n+1}h_{n+1}$, and $\left(
p-iq\right) h_{n}=\sqrt{n}h_{n-1}$, for $n=0,1,2,\dots $. As a result, the
respective matrices for $p$ and $q$ are as follows: 
\[
\frac{1}{2}\left( 
\begin{array}{cccccccc}
0 & 1 &  &  &  &  &  &  \\ 
1 & 0 & \sqrt{2} &  &  &  & \llap{\smash{\Huge $0$}} &  \\ 
& \sqrt{2} & 0 & \sqrt{3} &  &  &  &  \\ 
&  & \sqrt{3} & 0 & \ddots &  &  &  \\ 
&  &  & \ddots & \ddots & \ddots &  &  \\ 
&  &  &  & \ddots & 0 & \sqrt{n} &  \\ 
& \llap{\smash{\Huge $0$}} &  &  &  & \sqrt{n} & 0 & \ddots \\ 
&  &  &  &  &  & \ddots & \ddots
\end{array}
\right) 
\]
and 
\[
\frac{1}{2\sqrt{-1}}\left( 
\begin{array}{cccccccc}
0 & -1 &  &  &  &  &  &  \\ 
1 & 0 & -\sqrt{2} &  &  &  & \llap{\smash{\Huge $0$}} &  \\ 
& \sqrt{2} & 0 & -\sqrt{3} &  &  &  &  \\ 
&  & \sqrt{3} & 0 & \ddots &  &  &  \\ 
&  &  & \ddots & \ddots & \ddots &  &  \\ 
&  &  &  & \ddots & 0 & -\sqrt{n} &  \\ 
& \llap{\smash{\Huge $0$}} &  &  &  & \sqrt{n} & 0 & \ddots \\ 
&  &  &  &  &  & \ddots & \ddots
\end{array}
\right) . 
\]

One reason for the more general formulation (\ref{eqInt.5}) below, which is
based on an increasing resolution of orthogonal projections, as opposed to
an orthonormal basis, is that techniques which are effective in the simplest
case of tridiagonal matrices carry over to the case when some symmetric
operator $S$ is formed by taking noncommuting functions of the $p$, $q$
variables. Some such functions $S$ are known to be essentially selfadjoint,
and others not: for example, every quadratic expression $S_{2}$ in $p$, $q$
which is symmetric is essentially selfadjoint, while $S_{3}=pqp$ is not.
Also $S_{4}=p^{2}-q^{4}$ is not essentially selfadjoint, but $p^{2}+q^{4}$
is. These questions are also appropriate when we have instead an infinite
number of degrees of freedom, i.e., infinite systems $\left\{
p_{i},q_{j}\right\} $; see, e.g., Refs.\ \CITE{Far79,Pow74}.

In yet other applications, an orthonormal basis may not be readily
available, while a resolution of projections $\left\{ P_{n}\right\} $ may
be. For example, such a resolution may come from a wavelet construction: in
this case, usually such a resolution of projections is given at the outset,
while the wavelet basis is technically much more complicated.

This classical setup from (\ref{eqInt.1}) serves as motivation for the
results in the present paper. We now formulate a related geometric Hilbert
space problem which turns out to generalize the one above, and which, as
noted in Refs.\ \CITE{Jor75,Jor76,Jor77,Wer90} and above, has many more
applications.

\begin{definition}
\label{DefInt.1}\upshape An operator $S$ with dense domain ${\cal D}\left(
S\right) $ in a Hilbert space ${\cal H}$ is said to be {\em symmetric }(or
Hermitian) if 
\begin{equation}
\left\langle Sx,y\right\rangle =\left\langle x,Sy\right\rangle \text{\qquad
for all }x,y\in {\cal D}\left( S\right) .  \label{eqInt.3}
\end{equation}
In terms of the adjoint operator $S^{\ast }$, this means that $S\subset
S^{\ast }$, or equivalently, the domain of $S$ is contained in that of $%
S^{\ast }$, and $Sx=S^{\ast }x$ for $x\in {\cal D}\left( S\right) $.
\end{definition}

It was proved by von Neumann\cite{DS2} that the closure $\bar{S}$ is
selfadjoint in ${\cal H}$ if and only if both of the {\em deficiency spaces} 
\begin{equation}
{\cal E}_{\pm }\left( S\right) =\left\{ x_{\pm }\in {\cal D}\left( S^{\ast
}\right) \mathrel{;}S^{\ast }x_{\pm }=\pm ix_{\pm }\right\}  \label{eqInt.4}
\end{equation}
are zero. We say that a symmetric operator $S$ is {\em smooth} if there is a
sequence of projections 
\begin{equation}
P_{1}\leq P_{2}\leq \cdots  \label{eqInt.5}
\end{equation}
such that $P_{j}{\cal H}\subset {\cal D}\left( S\right) $, and $%
\bigcup_{j}P_{j}{\cal H}$ is dense in ${\cal H}$. The last density condition
may be restated as 
\begin{equation}
\sup_{j}P_{j}=I  \label{eqInt.6}
\end{equation}
where $I$ denotes the identity operator in ${\cal H}$. As noted by Stone, 
\cite{Sto32} if $S$ has dense domain, such projections $P_{j}$ may always be
constructed: we may even construct some smooth $\left\{ P_{j}\right\} $
which has each $P_{j}$ finite-dimensional, and for a given choice, there may
be others which have better estimates on the corresponding off-diagonal
terms. It is clear that the setting of smooth symmetric operators
generalizes the classical matrix problem from (\ref{eqInt.1}). But for the
more general operators, the {\em deficiency indices} 
\begin{equation}
n_{\pm }:=\dim {\cal E}_{\pm }\left( S\right)  \label{eqInt.7}
\end{equation}
need not be equal, and they may be more than one on either side, or
infinite. For examples of this, see Ref.\ \CITE{Jor77}. Recall that it is
known\cite{DS2} that $S$ has selfadjoint extensions in ${\cal H}$ if and
only if $n_{+}=n_{-}$.

If $P_{j}$ is a sequence of projections associated to a given symmetric
operator $S$, it is then clear from the geometric and general formulation
that the operators $\left( I-P_{j}\right) SP_{j}$ may be viewed as
off-diagonal terms, in a sense which naturally generalizes the off-diagonal
terms $b_{j}$ from the matrix (\ref{eqInt.1}) above. This is so since 
\begin{equation}
P_{j}^{\perp }:=I-P_{j}  \label{eqInt.8}
\end{equation}
is the projection onto the orthogonal complement of $P_{j}{\cal H}$. In an
earlier paper\cite{Jor75} we used the operator norms $\left\| P_{j}^{\perp
}SP_{j}^{{}}\right\| $ as a measure for the off-diagonal terms, but as noted
in Ref.\ \CITE{Jor75} this is a rough measure, and in the general case, the
best result of this kind amounts to the assertion that boundedness of $%
\left\| P_{j}^{\perp }SP_{j}^{{}}\right\| $ (as $j\rightarrow \infty $)
implies essential selfadjointness.

In this paper, we refine this result, by considering the vectors $%
P_{j}^{\perp }SP_{j}^{{}}x$ rather than the {\em operator norms} $\left\|
P_{j}^{\perp }SP_{j}^{{}}\right\| $. For one thing, the corresponding
``local results'' (i.e., depending on vectors) are more precise, and,
secondly, it is difficult in practice to compute operator norms. As noted in
Sakai's book,\cite{Sak91} the quantity $\left\| P_{j}^{\perp
}SP_{j}^{{}}x\right\| $ represents surface energy in statistical mechanics
models.

\section{A basic estimate}

\label{Bas}Let $S$ be a symmetric operator with dense domain ${\cal D}:=%
{\cal D}\left( S\right) $ in a Hilbert space ${\cal H}$, and let 
\begin{equation}
{\cal E}:={\cal E}_{+}\left( S\right) =\left\{ x\in {\cal D}\left( S^{\ast
}\right) \mathrel{;}S^{\ast }x=ix\right\} .  \label{eqBas.1}
\end{equation}

\begin{lemma}
\label{LemBas.1}Let $P$ be a smooth projection, and set $P^{\perp }:=I-P$.
Then we have the estimate 
\begin{equation}
\left\| P^{\perp }SPx\right\| \left\| P^{\perp }x\right\| \geq \left\|
Px\right\| ^{2}  \label{eqBas.2}
\end{equation}
for all $x\in {\cal E}$.
\end{lemma}

{\it Proof.} If $x\in {\cal E}$ is given, then $S^{\ast }x=ix$. Applying $P$
to both sides, and taking inner products, we get 
\begin{equation}
\left\langle Px,S^{\ast }x\right\rangle =i\left\| Px\right\| ^{2}.
\label{eqBas.3}
\end{equation}
But $\left\langle Px,S^{\ast }x\right\rangle =\left\langle
SPx,x\right\rangle =\left\langle PSPx,x\right\rangle +\left\langle P^{\perp
}SPx,x\right\rangle =\left\langle SPx,Px\right\rangle +\left\langle P^{\perp
}SPx,P^{\perp }x\right\rangle $, and $\left\langle SPx,Px\right\rangle
=\left\langle Px,SPx\right\rangle =\overline{\left\langle
SPx,Px\right\rangle }$. Hence 
\begin{equation}
%TCIMACRO{\limfunc{Im}}%
%BeginExpansion
\mathop{\rm Im}%
%EndExpansion
\left\langle P^{\perp }SPx,P^{\perp }x\right\rangle =\left\| Px\right\| ^{2}.
\label{eqBas.4}
\end{equation}
Since $\left| 
%TCIMACRO{\limfunc{Im}}%
%BeginExpansion
\mathop{\rm Im}%
%EndExpansion
\left\langle P^{\perp }SPx,P^{\perp }x\right\rangle \right| \leq \left|
\left\langle P^{\perp }SPx,P^{\perp }x\right\rangle \right| \leq \left\|
P^{\perp }SPx\right\| \left\| P^{\perp }x\right\| $, the result of the lemma
follows.\quad $\square $

Since the projection $P$ has its range contained in ${\cal D}\left( S\right) 
$, the operator $SP$ is bounded even though $S$ is typically unbounded. The
norm of the operator $P^{\perp }SP$ is a rough measure of the defect from
selfadjointness for the given symmetric operator $S$, and we have the
following:

\begin{corollary}
\label{CorBas.2}Let $P$ be a smooth projection, and let $x\in {\cal E}$ {\rm %
(}the defect space for a given symmetric operator $S${\rm ).} Then we have
the estimate, 
\begin{equation}
\left\| P^{\perp }SP|_{{\cal E}}\right\| \left\| P^{\perp }x\right\| \geq
\left\| Px\right\| .  \label{eqBas.5}
\end{equation}
\end{corollary}

{\it Proof.} The result follows from the lemma, and the observation 
\begin{equation}
\left\| P^{\perp }SPx\right\| \leq \left\| P^{\perp }SP|_{{\cal E}}\right\|
\left\| Px\right\| .  \label{eqBas.6}
\end{equation}
If this is substituted into (\ref{eqBas.4}), and the term $\left\|
Px\right\| $ is divided out, (\ref{eqBas.5}) follows.\quad $\square $

\begin{corollary}
\label{CorBas.3}{\rm (Jorgensen\cite{Jor76})} Let $S$ be a symmetric
operator, and let $\left\{ P_{i}\right\} _{i=1}^{\infty }$ be a sequence of
smooth projections such that $\sup_{i}P_{i}=I$. Then if 
\begin{equation}
\sup_{i}\left\| P_{i}^{\perp }SP_{i}^{{}}\right\| <\infty ,  \label{eqBas.7}
\end{equation}
then $S$ is essentially selfadjoint.
\end{corollary}

{\it Proof.} Apply the estimate (\ref{eqBas.7}) to both of the deficiency
spaces 
\begin{equation}
{\cal E}_{\pm }=\left\{ x_{\pm }\in {\cal D}\left( S^{\ast }\right) %
\mathrel{;}S^{\ast }x_{\pm }=\pm ix_{\pm }\right\} .  \label{eqBas.8}
\end{equation}
The boundedness of$\left\{ \left\| P_{j}^{\perp }SP_{j}^{{}}\right\|
\right\} _{j=1}^{\infty }$ implies that 
\begin{equation}
\lim_{j\rightarrow \infty }\left\| P_{j}^{\perp }SP_{j}^{{}}\right\| \left\|
P_{j}^{\perp }x_{\pm }^{{}}\right\| =0  \label{eqBas.9}
\end{equation}
since $\left\| x_{\pm }\right\| =\lim_{j\rightarrow \infty }\left\|
P_{j}x_{\pm }\right\| $, and $\left\| P_{j}^{\perp }x_{\pm }^{{}}\right\|
^{2}=\left\| x_{\pm }\right\| ^{2}-\left\| P_{j}x_{\pm }\right\|
^{2}\rightarrow 0$. An application of (\ref{eqBas.5}) then yields $x_{\pm
}=0 $. This applies for any vector $x_{\pm }$ in either of the two
deficiency spaces ${\cal E}_{\pm }$. Hence both of these spaces must vanish,
and it follows that $S$ is essentially selfadjoint by von Neumann's theorem;
see, e.g., Ref.\ \CITE{DS2}.\quad $\square $

The conclusion is that nontrivial defect for the given symmetric operator $S$
implies unboundedness of the sequence of norms from (\ref{eqBas.7}).

\begin{corollary}
\label{CorBas.4}Let $S$ and $\left\{ P_{j}\right\} _{j=1}^{\infty }$ be as
in Corollary {\rm \ref{CorBas.3},} but assume further that 
\begin{equation}
\left( I-P_{j+1}\right) SP_{j}=0\text{\qquad for all }j=1,2,\dots .
\label{eqBas.10}
\end{equation}
Then, if one of the defect spaces ${\cal E}_{\pm }$ is nonzero, it follows
that every positive sequence $b_{j}$ such that the estimates 
\begin{equation}
\left\| P_{j}^{\perp }SP_{j}^{{}}\right\| \leq b_{j}  \label{eqBas.11}
\end{equation}
hold will satisfy 
\begin{equation}
\sum_{j=1}^{\infty }\frac{1}{b_{j}^{2}}<\infty .  \label{eqBas.12}
\end{equation}
\end{corollary}

{\it Proof.} Suppose for simplicity that $x\in {\cal E}_{+}$ and $x\neq 0$.
Then the added restriction (\ref{eqBas.10}) placed on the projections may be
incorporated into the argument as follows: From (\ref{eqBas.4}) we have 
\[
%TCIMACRO{\limfunc{Im}}%
%BeginExpansion
\mathop{\rm Im}%
%EndExpansion
\left\langle P_{j}^{\perp }SP_{j}^{{}}x,x\right\rangle =\left\|
P_{j}x\right\| ^{2}. 
\]
But 
\[
\left\langle P_{j}^{\perp }SP_{j}^{{}}x,x\right\rangle =\left\langle
P_{j}^{\perp }SP_{j}^{{}}x,\left( P_{j+1}-P_{j}\right) x\right\rangle 
\]
if (\ref{eqBas.10}) is assumed. Hence 
\[
\left\| P_{j}x\right\| ^{2}\leq b_{j}\left\| P_{j}x\right\| \left\|
P_{j+1}x-P_{j}x\right\| , 
\]
where the estimates (\ref{eqBas.11}) are used. Cancelling a $\left\|
P_{j}x\right\| $ factor, we get 
\[
\left\| P_{j}x\right\| \leq b_{j}\left( \left\| P_{j+1}x\right\|
^{2}-\left\| P_{j}x\right\| ^{2}\right) ^{1/2}. 
\]
Equivalently, with the estimates (\ref{eqBas.11}), we therefore get 
\[
b_{j}^{2}\left\| \left( P_{j+1}-P_{j}\right) x\right\| ^{2}\geq \left\|
P_{j}x\right\| ^{2} 
\]
and 
\[
b_{j}^{2}\left\| P_{j+1}x\right\| ^{2}\geq \left( 1+b_{j}^{2}\right) \left\|
P_{j}x\right\| ^{2}. 
\]
Hence 
\[
\left\| P_{k+j}x\right\| ^{2}\geq \prod_{s=j}^{k+k-1}\left( 1+\frac{1}{%
b_{s}^{2}}\right) \left\| P_{j}x\right\| ^{2}. 
\]
If $x\neq 0$, then there is some $j$ such that $P_{j}x\neq 0$ and so the
product $\prod_{j\leq x<j+k}\left( 1+1/b_{s}^{2}\right) $ converges as $%
k\rightarrow \infty $, which implies the finiteness of the sum (\ref
{eqBas.12}), and the result follows.\quad $\square $

The following result is stronger than the corollary:

\begin{theorem}
\label{ThmBas.5}With the assumptions of Corollary {\rm \ref{CorBas.4},} and
assuming further that one of the deficiency spaces ${\cal E}_{\pm }$ is
nonzero, we get $\sum_{j=1}^{\infty }b_{j}^{-1}<\infty $.
\end{theorem}

{\it Proof.} As in the previous proof, let $x\in {\cal E}_{+}\setminus
\left\{ 0\right\} $. By virtue of (\ref{eqBas.10}) we have 
\[
\left\langle P_{j}^{\perp }SP_{j}^{{}}x,x\right\rangle =\left\langle
P_{j}^{\perp }SP_{j}^{{}}\left( P_{j}-P_{j-1}\right) x,\left(
P_{j+1}-P_{j}\right) x\right\rangle 
\]
and therefore 
\[
\left\| P_{j}x\right\| ^{2}\leq b_{j}\left\| \left( P_{j}-P_{j-1}\right)
x\right\| \left\| \left( P_{j+1}-P_{j}\right) x\right\| . 
\]
Pick $j$ such that $P_{j}x\neq 0$. We will show that the numbers $\left(
b_{k}\right) $ from (\ref{eqBas.11}) yield $\left( b_{k}^{-1}\right)
_{k>j}\in \ell ^{1}$. Summing the previous estimate, we now get the
following: 
\begin{eqnarray*}
\left\| P_{j}x\right\| ^{2}\sum_{k=j}^{n}b_{k}^{-1} &\leq &\left(
\sum_{k=j}^{n}\left\| \left( P_{k}-P_{k-1}\right) x\right\|
^{2}\sum_{k=j+1}^{n+1}\left\| \left( P_{k}-P_{k-1}\right) x\right\|
^{2}\right) ^{1/2} \\
&=&\left\| \left( P_{n}-P_{j-1}\right) x\right\| \left\| \left(
P_{n+1}-P_{j}\right) x\right\| \\
&\leq &\left\| x\right\| ^{2},
\end{eqnarray*}
and the result follows.\quad $\square $

\section{Local estimates}

\label{Loc}The norm estimates on $P_{j}^{\perp }SP_{j}^{{}}$ in the previous
section are rather rough in that they do not yield direct asymptotic
properties of the sequences $P_{j}^{\perp }SP_{j}^{{}}x$ for fixed vectors $%
x $, and we now turn to this question, beginning with a basic lemma:

\begin{lemma}
\label{LemLoc.1}Let $S$ be a symmetric operator as in Section {\rm \ref{Bas},%
} and let $x$ be a nonzero vector in one of the deficiency spaces {\rm (\ref
{eqBas.8}).} Then 
\begin{equation}
\lim_{j\rightarrow \infty }\left\| P_{j}^{\perp }SP_{j}^{{}}x\right\|
=\infty ;  \label{eqLoc.1}
\end{equation}
more specifically, $\left\| P_{j}x\right\| <\left\| x\right\| $ for all $j$,
and 
\begin{equation}
\left\| P_{j}^{\perp }SP_{j}^{{}}x\right\| \geq \frac{\left\| P_{j}x\right\|
^{2}}{\sqrt{\left\| x\right\| ^{2}-\left\| P_{j}x\right\| ^{2}}}.
\label{eqLoc.2}
\end{equation}
\end{lemma}

{\it Proof.} From (\ref{eqBas.4}) in Section \ref{Bas}, we get 
\begin{equation}
\left\| P_{j}^{\perp }SP_{j}^{{}}x\right\| \left\| P_{j}^{\perp }x\right\|
\geq \left\| P_{j}x\right\| ^{2}.  \label{eqLoc.3}
\end{equation}
Since $\left\| P_{j}^{\perp }x\right\| =\sqrt{\left\| x\right\| ^{2}-\left\|
P_{j}x\right\| ^{2}}$, the estimate (\ref{eqLoc.2}) follows. The conclusion (%
\ref{eqLoc.1}) is implied by the estimate since $\lim_{j\rightarrow \infty
}\left\| P_{j}x\right\| ^{2}=\left\| x\right\| ^{2}$ from the assumption (%
\ref{eqInt.6}) on the sequence $P_{j}$. We always have $\left\|
P_{j}x\right\| \leq \left\| x\right\| $, but this inequality is sharp. For
if $\left\| P_{j}x\right\| =\left\| x\right\| $, then $P_{j}x=x$, and this
contradicts the fact 
\begin{equation}
{\cal D}\left( S\right) \cap {\cal E}_{\pm }=\left\{ 0\right\} .
\label{eqLoc.4}
\end{equation}
To see this, note that $P_{j}x\in {\cal D}\left( S\right) $ since $P_{j}$ is
smooth, and $x\in {\cal E}_{+}$ by assumption. The conclusion of the lemma
follows.\quad $\square $

We now turn to the more restrictive class of symmetric operators $S$
considered in Corollary \ref{CorBas.4}: While we have the subspaces ${\cal D}%
_{j}=P_{j}{\cal H}$ contained in the domain of $S$ for all $j$, the more
restrictive assumption in Corollary \ref{CorBas.4} is that $S$ maps ${\cal D}%
_{j}$ into the next space ${\cal D}_{j+1}$. Note that, if $\bigcup_{j}{\cal D%
}_{j}$ is mapped into itself, we can always arrange, by relabeling the
indexing of the subspaces, that this condition is satisfied relative to the
relabeled sequence of subspaces. This means that the corresponding
projections $P_{j}$ will then satisfy the property (\ref{eqBas.10}) from
Corollary \ref{CorBas.4}. We now turn to the sequence $\left\| P_{j}^{\perp
}SP_{j}^{{}}x\right\| $ for the case when $S$ has at least one nontrivial
deficiency space, i.e., when one of the two spaces ${\cal E}_{\pm }\left(
S\right) $ is assumed nonzero. We already showed in Lemma \ref{LemLoc.1}
that then $\left\| P_{j}^{\perp }SP_{j}^{{}}x\right\| \rightarrow \infty $
as $j\rightarrow \infty $, and so we may assume without loss of generality
that the numbers $\left\| P_{j}^{\perp }SP_{j}^{{}}x\right\| $ are all
nonzero.

The next result specifies a growth rate for this sequence in case of nonzero
deficiency. To apply the result in proving essential selfadjointness of some
given symmetric operator $S$, with the scaling property (\ref{eqLoc.6}), we
can then check that the sequence $\left\| P_{j}^{\perp }SP_{j}^{{}}x\right\| 
$ grows less rapidly than the {\em a priori} rate (see (\ref{eqLoc.8})) and $%
S$ must then have selfadjoint closure.

\begin{theorem}
\label{ThmLoc.2}Let $S$ be a densely defined symmetric operator which has a
smooth sequence of projections $P_{j}$ with 
\begin{equation}
\sup_{j}P_{j}=I  \label{eqLoc.5}
\end{equation}
and 
\begin{equation}
P_{j+1}SP_{j}=SP_{j}\text{\qquad for all }j=1,2,\dots .  \label{eqLoc.6}
\end{equation}
Suppose one of the deficiency spaces is nonzero. Let $x\in {\cal E}\setminus
\left\{ 0\right\} $. Then there is a number $\xi \in \left( 0,1\right) $,
the open unit interval, such that the local sequence 
\begin{equation}
c_{j}=c_{j}\left( x\right) :=\left\| P_{j}^{\perp }SP_{j}^{{}}x\right\| ^{2}
\label{eqLoc.7}
\end{equation}
grows so rapidly as to yield existence of the limit 
\begin{equation}
\lim_{j\rightarrow \infty }F_{c_{j+1}}\left( F_{c_{j}}\left( \cdots \left(
F_{c_{2}}\left( F_{c_{1}}\left( \xi \right) \right) \right) \cdots \right)
\right) \leq 1,  \label{eqLoc.8}
\end{equation}
where 
\begin{equation}
F_{c}\left( s\right) =s+\frac{1}{c}s^{2},\qquad s\in {\Bbb R}_{+}.
\label{eqLoc.9}
\end{equation}
Moreover, then $\sum_{j}1/c_{j}<\infty $.
\end{theorem}

{\it Proof.} The start of the proof is the same as that of Corollary \ref
{CorBas.4}, but then the next estimate is refined as follows: Suppose for
specificity that $x\in {\cal E}_{+}$, and that $\left\| x\right\| =1$. By
virtue of the assumption on $S$, we have 
\begin{equation}
\left\langle P_{j}^{\perp }SP_{j}^{{}}x,x\right\rangle =\left\langle
P_{j}^{\perp }SP_{j}^{{}}x,\left( P_{j+1}-P_{j}\right) x\right\rangle
\label{eqLoc.10}
\end{equation}
and 
\[
%TCIMACRO{\limfunc{Im}}%
%BeginExpansion
\mathop{\rm Im}%
%EndExpansion
\left\langle P_{j}^{\perp }SP_{j}^{{}}x,x\right\rangle =\left\|
P_{j}x\right\| ^{2}. 
\]
As a consequence, we get the estimate 
\begin{equation}
\left\| P_{j}x\right\| ^{2}\leq c_{j}^{1/2}\left\| P_{j+1}x-P_{j}x\right\| ,
\label{eqLoc.11}
\end{equation}
where the sequence $c_{j}$ is defined in (\ref{eqLoc.7}). Introducing $\xi
_{j}:=\left\| P_{j}x\right\| ^{2}$, we note that $0<\xi _{j}<1$, and $%
\lim_{j\rightarrow \infty }\xi _{j}=1$ ($=\left\| x\right\| ^{2}$), and the
limit is monotone. Also note that the functions $F_{c}$ from (\ref{eqLoc.9})
are monotone on ${\Bbb R}_{+}$. Now the estimate (\ref{eqLoc.11}) takes the
form 
\begin{equation}
\xi _{j}^{2}\leq c_{j}\left( \xi _{j+1}-\xi _{j}\right) ,  \label{eqLoc.12}
\end{equation}
or equivalently 
\begin{equation}
F_{c_{j}}\left( \xi _{j}\right) \leq \xi _{j+1}\text{\qquad for all }%
j=1,2,\dots .  \label{eqLoc.13}
\end{equation}
Starting with the initial estimate $F_{c_{1}}\left( \xi _{1}\right) \leq \xi
_{2}$, and using monotonicity of $F_{c_{2}}$, we get 
\[
F_{c_{2}}\left( F_{c_{1}}\left( \xi _{1}\right) \right) \leq F_{c_{2}}\left(
\xi _{2}\right) \leq \xi _{3}, 
\]
and then, by induction, 
\[
%TCIMACRO{
%\underset{=:t_{j}}{\underbrace{F_{c_{j}}\left( F_{c_{j-1}}\left( \cdots \left( F_{c_{2}}\left( F_{c_{1}}\left( \xi _{1}\right) \right) \right) \cdots \right) \right) }}}%
%BeginExpansion
\mathrel{\mathop{\underbrace{F_{c_{j}}\left( F_{c_{j-1}}\left( \cdots \left( F_{c_{2}}\left( F_{c_{1}}\left( \xi _{1}\right) \right) \right) \cdots \right) \right) }}\limits_{=:t_{j}}}%
%EndExpansion
\leq \xi _{j+1}. 
\]
It also follows from (\ref{eqLoc.9}) that the sequence $t_{j}$ is monotone
and strictly increasing. Indeed, $t_{j+1}=t_{j}+\frac{1}{c_{j+1}}%
t_{j}^{2}>t_{j}$ for all $j$. Since $\xi _{j+1}=\left\| P_{j+1}x\right\|
^{2}<1$, with limit $=1$, the result follows. The last conclusion in the
theorem will be proved in the next section.\quad $\square $

\begin{remark}
\label{RemLoc.3}\upshape It follows from the theorem that if the containment 
$\bar{S}\subset S^{\ast }$ is strict, then there are vectors $x$ in ${\cal D}%
\left( S^{\ast }\right) $ such that the boundary terms $c_{n}\left( x\right)
=\left\| P_{n}^{\perp }SP_{n}^{{}}x\right\| ^{2}$ have growth asymptotics at
least some {\em a priori} rate. If further (\ref{eqLoc.6}) is assumed, then
this rate may be specified as 
\begin{equation}
\sum_{n}c_{n}\left( x\right) ^{-1}<\infty .  \label{eqLoc.14}
\end{equation}
But we have the following converse: Suppose, for some $x\in {\cal D}\left(
S^{\ast }\right) $, that (\ref{eqLoc.14}) holds. Then $x$ cannot be in the
domain of $\bar{S}$, and so $\bar{S}\subsetneqq S^{\ast }$, or equivalently, 
$S$ is then not essentially selfadjoint. This is strong enough for proving
that the noncommutative polynomials $pqp$ and $p^{2}-q^{4}$ from Section \ref
{Int} are not essentially selfadjoint on the span ${\cal D}$ of the Hermite
polynomials, and therefore {\em a fortiori} also not on ${\cal S}$.

To prove the claim, suppose a symmetric operator $S$ is given to satisfy the
conditions in Theorem \ref{ThmLoc.2}. We claim that, if $x\in {\cal D}\left( 
\bar{S}\right) $, then $c_{n}\left( x\right) =\left\| P_{n}^{\perp
}SP_{n}^{{}}x\right\| ^{2}$ is bounded. This clearly is inconsistent with (%
\ref{eqLoc.14}), so if (\ref{eqLoc.14}) holds for some $x\in {\cal D}\left(
S^{\ast }\right) $, then $\bar{S}$ is not selfadjoint. We now prove
boundedness of $c_{n}\left( x\right) $ for $x\in {\cal D}\left( \bar{S}%
\right) $: If $x\in {\cal D}\left( \bar{S}\right) $, then $%
SP_{n}x\rightarrow \bar{S}x$, and so it is enough to prove boundedness if $%
x\in {\cal D}:=\bigcup_{n}P_{n}{\cal H}$ ($\subset {\cal D}\left( S\right) $%
). Let $x\in P_{n}{\cal H}$, i.e., $P_{n}x=x$. Then $%
P_{j}x=P_{j}P_{n}x=P_{n}x$ for all $j\geq n$, and, using (\ref{eqLoc.6}), $%
P_{j}^{\perp }SP_{j}^{{}}x=P_{j}^{\perp }SP_{n}^{{}}x=P_{j}^{\perp
}P_{n+1}^{{}}Sx=0$ if $j\geq n+1$. Hence $c_{j}\left( x\right) =0$ if $j\geq
n+1$, and $\lim_{j\rightarrow \infty }c_{j}\left( x\right) =0$ for all $x\in 
{\cal D}$. We leave the remaining approximation argument for the reader,
i.e., passing to vectors $x$ in ${\cal D}\left( \bar{S}\right) $.
\end{remark}

\section{Functional iteration}

\label{Fun}The condition (\ref{eqLoc.8}) of Theorem \ref{ThmLoc.2} is
perhaps not as transparent as the corresponding condition (\ref{eqBas.12})
in Corollary \ref{CorBas.4}. But there is a simple comparison between the
two sequences 
\begin{equation}
b_{j}=\left\| P_{j}^{\perp }SP_{j}^{{}}\right\| \text{\quad and\quad }%
c_{j}\left( x\right) =\left\| P_{j}^{\perp }SP_{j}^{{}}x\right\| ^{2}.
\label{eqFun.1}
\end{equation}
Clearly 
\begin{equation}
c_{j}\left( x\right) \leq \left\| x\right\| ^{2}b_{j}^{2}.  \label{eqFun.2}
\end{equation}
So if $x$ is a nonzero vector in one of the two deficiency spaces ${\cal E}%
_{\pm }$, then 
\begin{equation}
\sum_{j}\frac{1}{b_{j}^{2}}\leq \left\| x\right\| ^{2}\sum_{j}\frac{1}{%
c_{j}\left( x\right) }.  \label{eqFun.3}
\end{equation}
The condition from Corollary \ref{CorBas.4} is (\ref{eqBas.12}), and its
negation, 
\begin{equation}
\sum_{j}\frac{1}{b_{j}^{2}}=\infty ,  \label{eqFun.pound}
\end{equation}
implies selfadjointness. We conclude then that if $x$ is any nonzero vector,
then the condition 
\begin{equation}
\sum_{j}\frac{1}{c_{j}\left( x\right) }=\infty  \label{eqFun.4}
\end{equation}
follows from (\ref{eqFun.pound}). We now show that Theorem \ref{ThmLoc.2} is
strictly stronger then Corollary \ref{CorBas.4}. To this end we state a
simple lemma on functional iteration which explains the two types of
estimates involved.

\begin{lemma}
\label{LemFun.3}Let $\left\{ c_{j}\right\} _{j=1}^{\infty }\subset {\Bbb R}%
_{+}$ be given. Then the following two conditions are equivalent:

\begin{enumerate}
\item  \label{LemFun.3(1)}$\sum_{j}1/c_{j}<\infty $;

\item  \label{LemFun.3(2)}There is a $\xi \in \left( 0,1\right) $ such that 
\begin{equation}
F_{c_{j}}\left( F_{c_{j-1}}\left( \cdots \left( F_{c_{2}}\left(
F_{c_{1}}\left( \xi \right) \right) \right) \cdots \right) \right) <1\text{%
\qquad for all }j=1,2,\dots .  \label{eqFun.5}
\end{equation}
\end{enumerate}
\end{lemma}

{\it Proof.} (\ref{LemFun.3(1)}) $\Rightarrow $ (\ref{LemFun.3(2)}): Assume (%
\ref{LemFun.3(1)}). Since $\ln \left( 1+1/c_{i}\right) <1/c_{i}$, we get $%
\ln \prod_{i=1}^{j}\left( 1+1/c_{i}\right) <\sum_{i=1}^{j}1/c_{i}$, the
infinite product then converges, and $\prod_{i=1}^{\infty }\left(
1+1/c_{i}\right) <\exp \left( \sum_{i=1}^{\infty }1/c_{i}\right) <\infty $.
Hence we may pick $\xi \in \left( 0,1\right) $ such that 
\begin{equation}
\prod_{i=1}^{j}\left( 1+\frac{1}{c_{i}}\right) \xi <\prod_{i=1}^{\infty
}\left( 1+\frac{1}{c_{i}}\right) \xi <1\text{\qquad for all }j.
\label{eqFun.6}
\end{equation}
If we prove that 
\begin{equation}
t_{j}:=F_{c_{j}}\left( F_{c_{j-1}}\left( \cdots \left( F_{c_{1}}\left( \xi
\right) \right) \cdots \right) \right) <\prod_{i=1}^{j}\left( 1+\frac{1}{%
c_{i}}\right) \xi ,  \label{eqFun.7}
\end{equation}
then the first implication of the lemma follows. But this is an induction
argument: Suppose it holds up to $j-1$. Then by (\ref{eqFun.6}), we will
have $t_{j-1}=F_{c_{j-1}}\left( F_{c_{j-2}}\left( \cdots \left(
F_{c_{1}}\left( \xi \right) \right) \cdots \right) \right) <1$, and
therefore, using $t_{j-1}^{2}<t_{j-1}^{{}}$, we get 
\[
t_{j}=F_{c_{j}}\left( t_{j-1}\right) =t_{j-1}+\frac{1}{c_{j}}%
t_{j-1}^{2}<\left( 1+\frac{1}{c_{j}}\right) t_{j-1}<\prod_{i=1}^{j}\left( 1+%
\frac{1}{c_{i}}\right) \xi , 
\]
where, in the last step, the induction hypothesis was used a second time.
This concludes the induction step, and (\ref{eqFun.7}) is proved. By the
choice of $\xi $ in (\ref{eqFun.6}), we now get the desired conclusion (\ref
{eqFun.5}) of the lemma.\quad $\square $

(\ref{LemFun.3(2)}) $\Rightarrow $ (\ref{LemFun.3(1)}): Assume (\ref
{LemFun.3(2)}). The first two estimates in (\ref{eqFun.5}) are 
\[
t_{1}=\xi +\frac{1}{c_{1}}\xi ^{2}<1, 
\]
and 
\[
t_{2}=\xi +\frac{1}{c_{1}}\xi ^{2}+\frac{1}{c_{2}}\left( \xi +\frac{1}{c_{1}}%
\xi ^{2}\right) ^{2}<1. 
\]
Completing the second square, we get five positive terms in the sum on the
left, and so {\em a fortiori} 
\[
\frac{1}{c_{1}}\xi ^{2}+\frac{1}{c_{2}}\xi ^{2}<1 
\]
when only two out of the five terms are retained in the sum. But the general
term 
\[
t_{j}=F_{c_{j}}\left( F_{c_{j-1}}\left( \cdots \left( F_{c_{2}}\left(
F_{c_{1}}\left( \xi \right) \right) \right) \cdots \right) \right) 
\]
on the left-hand side in (\ref{eqFun.5}) includes, when all the squares are
completed, the following $j$ terms: 
\[
\frac{1}{c_{1}}\xi ^{2}+\frac{1}{c_{2}}\xi ^{2}+\cdots +\frac{1}{c_{j}}\xi
^{2}\text{\qquad (}<t_{j}\text{)} 
\]
among a total of $3\cdot \left( j-1\right) $ positive terms. Since all these
terms sum up to $t_{j}<1$, we get, for the retained ones, $\left(
\sum_{i=1}^{j}1/c_{i}\right) \xi ^{2}<t_{j}<1$, and therefore $%
\sum_{i=1}^{\infty }1/c_{i}<\infty $.\quad $\square $

\begin{proposition}
\label{ProFun.4}The implication in Corollary {\rm \ref{CorBas.4}} may be
derived from that of Theorem {\rm \ref{ThmLoc.2},} and Theorem {\rm \ref
{ThmLoc.2}} applies to cases not covered by Corollary {\rm \ref{CorBas.4}.}
\end{proposition}

{\it Proof.} Suppose $x\in {\cal E}\left( S\right) \setminus \left\{
0\right\} $. Normalize such that $\left\| x\right\| =1$. Then, by Theorem 
\ref{ThmLoc.2}, we may pick $j$ such that $\xi =\left\| P_{j}x\right\|
^{2}\in \left( 0,1\right) $ will satisfy $F_{c_{j+k}}\left(
F_{c_{j+k-1}}\left( \cdots \left( F_{c_{j+1}}\left( \xi \right) \right)
\cdots \right) \right) <1$ for all $k$. Using the lemma, we get $%
\sum_{j}1/c_{j}<\infty $, and by (\ref{eqFun.3}), we must then have $%
\sum_{j}1/b_{j}^{2}<\infty $. This shows that Theorem \ref{ThmLoc.2} is the
stronger of the two results. The examples in Ref.\ \CITE{Jor77} further show
that, in fact, the result in Section \ref{Loc} is strictly stronger than
that of Section \ref{Bas}.\quad $\square $

\section{Positive operators}

\label{Pos}We say that a symmetric operator $L$ with dense domain in a
Hilbert space ${\cal H}$ is positive if 
\begin{equation}
\left\langle y,Ly\right\rangle \geq 0,\qquad y\in {\cal D}\left( L\right) .
\label{eqPos.1}
\end{equation}
For vectors $x\in {\cal H}$, then the sequence 
\begin{equation}
d_{n}\left( x\right) :=\left\langle x,P_{n}^{\perp
}LP_{n}^{{}}x\right\rangle ,\qquad n\in {\Bbb N},  \label{eqPos.2}
\end{equation}
is a more natural measure for off-diagonal terms, relative to some system $%
\left\{ P_{n}\right\} _{n=1}^{\infty }$ of smooth projections with $%
\sup_{n}P_{n}=I$; and we have the following obvious estimate: 
\begin{equation}
d_{n}\left( x\right) \leq c_{n}\left( x\right) ^{1/2}\left\| x\right\| ,
\label{eqPos.3}
\end{equation}
where $c_{n}\left( x\right) =c_{n}\left( L,x\right) :=\left\| P_{n}^{\perp
}LP_{n}^{{}}x\right\| ^{2}$. As a result, we have the following estimate for
the sums 
\begin{equation}
\sum_{n}c_{n}\left( x\right) ^{-1/2}\leq \left\| x\right\|
\sum_{n}d_{n}\left( x\right) ^{-1}  \label{eqPos.4}
\end{equation}
for all nonzero vectors $x$ in ${\cal H}$. Hence, if a given symmetric
operator $L$ is also known to be positive, then we get the following
improvement on Theorem \ref{ThmLoc.2}.

\begin{theorem}
\label{ThmPos.1}Let $L$ be a positive operator, and suppose $\left\{
P_{n}\right\} _{n=1}^{\infty }$ are smooth projections satisfying 
\begin{equation}
\sup_{n}P_{n}=I,  \label{eqPos.5}
\end{equation}
and suppose further, for some $k\in {\Bbb N}$, that 
\begin{equation}
P_{n+k}LP_{n}=LP_{n}\text{\qquad for all }n.  \label{eqPos.6}
\end{equation}
If $x\in {\cal E}_{\pm }\left( L\right) \setminus \left\{ 0\right\} $, then $%
\sum_{n}c_{n}\left( x\right) ^{-1/2}<\infty $.
\end{theorem}

{\it Proof.} The result in fact is a consequence of the following more
general one, combined with (\ref{eqPos.4}).\quad $\square $

\begin{theorem}
\label{ThmPos.2}Let $L$ be a positive operator, and let $\left\{
P_{n}\right\} _{n=1}^{\infty }$ be smooth projections satisfying {\rm (\ref
{eqPos.5})--(\ref{eqPos.6}).} Then, if $x\in {\cal D}\left( L^{\ast }\right)
\setminus \left\{ 0\right\} $ satisfies $L^{\ast }x=-x$, we get the
summability: 
\begin{equation}
\sum_{n}d_{n}\left( x\right) ^{-1}<\infty .  \label{eqPos.7}
\end{equation}
\end{theorem}

{\it Proof.} This result is implicit in the proof of Lemma 1 in Ref.\ \cite
{Jor78}. We will also need the general fact\cite{DS2} that positive
operators $L$ with dense domain are essentially selfadjoint if and only if $%
\left\{ x\in {\cal D}\left( L^{\ast }\right) \mathrel{;}L^{\ast
}x=-x\right\} =\left\{ 0\right\} $.\quad $\square $

Let $\left\{ S_{i}\right\} _{i=1}^{k}$ be a finite family of symmetric
operators in a Hilbert space ${\cal H}$ which are defined on a common dense
invariant domain ${\cal D}$ in ${\cal H}$. Then $L:=\sum_{i=1}^{k}S_{i}^{2}$
is positive and defined on ${\cal D}$. Nelson\cite{Nel59b} and Poulsen\cite
{Pou73} studied the question of deciding when the operators $\bar{S}_{i}$
are selfadjoint with commuting spectral resolutions. A necessary condition
for this is the commutativity 
\begin{equation}
S_{i}S_{j}y=S_{j}S_{i}y\text{\qquad for all }i,j\leq k\text{, and all }y\in 
{\cal D}.  \label{eqPos.8}
\end{equation}
If such commuting spectral resolutions exist, then there is, by Refs.\ %
\CITE{DS2,AkGl93}, a spectral measure $E\left( \,\cdot \,\right) $ on ${\Bbb %
R}^{k}$ such that 
\begin{equation}
S_{i}x=\int_{{\Bbb R}^{k}}\lambda _{i}\,dE\left( \lambda \right) x
\label{eqPos.9}
\end{equation}
and 
\begin{equation}
\left\| x\right\| ^{2}=\int_{{\Bbb R}^{k}}\left\| dE\left( \lambda \right)
x\right\| ^{2},  \label{eqPos.10}
\end{equation}
where we use the standard notation ${\Bbb R}^{k}\ni \lambda =\left( \lambda
_{1},\dots ,\lambda _{k}\right) $.

Nelson's celebrated theorem (see also Ref.\ \CITE{Pou73}) states that a
joint spectral resolution (\ref{eqPos.9}) exists if the $S_{i}$'s satisfy (%
\ref{eqPos.8}), and if $L=\sum_{i}S_{i}^{2}$ is essentially selfadjoint on $%
{\cal D}$.

Our off-diagonal terms from Theorem \ref{ThmLoc.2} are especially useful in
the multivariable case, as is illustrated in the following theorem.

\begin{theorem}
\label{ThmPos.3}Let $\left\{ S_{i}\right\} _{i=1}^{k}$ be given symmetric
operators satisfying {\rm (\ref{eqPos.8}).} Let $\left\{ P_{n}\right\}
_{n=1}^{\infty }$ be smooth projections such that $P_{n}{\cal H}\subset 
{\cal D}$, $\sup_{n}P_{n}=I$, and 
\begin{equation}
P_{n+1}S_{i}P_{n}=S_{i}P_{n}\text{\qquad for all }1\leq i\leq k\text{, and
all }n\in {\Bbb N},  \label{eqPos.11}
\end{equation}
i.e., each $S_{i}$ satisfying the conditions in Theorem {\rm \ref{ThmLoc.2}.}
Suppose, for all $x\in {\cal H}$, that we have the following asymptotics: 
\begin{equation}
c_{i}\left( n,x\right) =\left\| P_{n}^{\perp }S_{i}^{{}}P_{n}^{{}}x\right\|
^{2}\leq {\cal O}\left( n\right) .  \label{eqPos.12}
\end{equation}
Then the closed operators $\bar{S}_{i}$ are selfadjoint {\rm (}the $S_{i}$'s
are essentially selfadjoint on ${\cal D}${\rm ),} and they have joint
spectral resolution in the sense of {\rm (\ref{eqPos.9}).}
\end{theorem}

{\it Proof.} For Nelson's operator $L=\sum_{i=1}^{k}S_{i}^{2}$, we have
off-diagonal defect terms as follows: 
\begin{eqnarray*}
d\left( L,n,x\right) &=&\left\langle P_{n}^{\perp
}LP_{n}^{{}}x,x\right\rangle \\
&=&\left\langle L\left( P_{n}-P_{n-2}\right) x,\left( P_{n+2}-P_{n}\right)
x\right\rangle \\
&=&\sum_{i}\left\langle S_{i}\left( P_{n}-P_{n-2}\right) x,S_{i}\left(
P_{n+2}-P_{n}\right) x\right\rangle \\
&=&\sum_{i}\left\langle P_{n-1}^{\perp }S_{i}^{{}}P_{n}^{{}}x,P_{n+1}^{\perp
}S_{i}^{{}}P_{n+2}^{{}}x\right\rangle \\
&=&\sum_{i}\left\langle P_{n+1}^{\perp }S_{i}^{{}}P_{n}^{{}}x,P_{n+1}^{\perp
}S_{i}^{{}}P_{n+2}^{{}}x\right\rangle \\
&=&\sum_{i}\left\langle P_{n+1}^{\perp }P_{n}^{\perp
}S_{i}^{{}}P_{n}^{{}}x,P_{n+1}^{\perp }S_{i}^{{}}P_{n+1}^{{}}x\right\rangle
\\
&\leq &\sum_{i}\left\| P_{n}^{\perp }S_{i}^{{}}P_{n}^{{}}x\right\| \left\|
P_{n+1}^{\perp }S_{i}^{{}}P_{n+1}^{{}}x\right\| \\
&\leq &\left( \sum_{i=1}^{k}c_{i}\left( n,x\right) \sum_{j=1}^{k}c_{j}\left(
n+1,x\right) \right) ^{1/2} \\
&\leq &{\cal O}\left( n\right)
\end{eqnarray*}
by virtue of the assumption (\ref{eqPos.12}) made for each of the operators $%
S_{i}$. Each $S_{i}$ is essentially selfadjoint on ${\cal D}$ by Theorem \ref
{ThmLoc.2}; but {\em a priori} the corresponding spectral resolutions $%
E_{i}\left( \,\cdot \,\right) $ may not commute (see, e.g., Refs.\ %
\CITE{Nel59b}, \CITE{Vas99}, or \CITE{JoMu80,JoPo91}). However, since $%
d\left( L,n,x\right) \leq {\cal O}\left( n\right) $ for all $x\in {\cal H}$,
we conclude from Theorem \ref{ThmPos.2} above that $L=%
\sum_{i=1}^{k}S_{i}^{2} $ is then essentially selfadjoint on ${\cal D}$.
Hence, Nelson's theorem\cite{Nel59b} implies that the individual spectral
resolutions $E_{i}\left( \,\cdot \,\right) $ on ${\Bbb R}$ are mutually
commuting. So, if we define $E $ on ${\Bbb R}^{k}$ as a product measure, $%
dE\left( \lambda \right) =E_{1}\left( d\lambda _{1}\right) E_{2}\left(
d\lambda _{2}\right) \cdots E_{k}\left( d\lambda _{k}\right) $, then it
follows from standard spectral theory (see, e.g., Ref.\ \CITE{DS2}) that $%
E\left( \,\cdot \,\right) $ (on ${\Bbb R}^{k}$) will satisfy (\ref{eqPos.9}%
)--(\ref{eqPos.10}).\quad $\square $

\begin{remark}
\label{RemPos.4}\upshape The multivariable case in the present section is
especially useful in recent work on multivariable spectral theory by
Vasilescu et al.\cite{Vas99,PuVa99b} There, in applications to multivariable
moment problems, the issue of commutativity of symmetric operators in the
weak sense, versus the strong sense, is related to comparison of joint
distributions vs.\ marginal distributions.
\end{remark}

Other applications to mathematical physics are sketched in Refs.\ %
\CITE{JoPo91} and \CITE{Jor75,Jor76,Jor77,Jor78,Jor92b}. In particular, our
assumptions are especially useful in the study of noncommutative polynomials
applied to quantum fields like momentum and position bosonic variables, as
they are given traditionally in terms of raising and lowering operators.

A simple application of Theorem \ref{ThmPos.2} then yields the following
concrete corollary: Let 
\[
\left( p_{i}h\right) \left( x\right) =\frac{1}{\sqrt{-1}}\frac{\partial \;}{%
\partial x_{i}}\left( x\right) , 
\]
and 
\[
\left( q_{i}h\right) \left( x\right) =x_{j}h\left( x_{1},\dots ,x_{k}\right) 
\]
for $h\in {\cal S}\left( {\Bbb R}^{k}\right) \subset L^{2}\left( {\Bbb R}%
^{k}\right) $. Let $L$ be a noncommutative polynomial in the variables $%
p_{i} $, $q_{j}$ for $1\leq i,j\leq k$ of degree at most four, such that $%
\left\langle h,Lh\right\rangle _{L^{2}\left( {\Bbb R}^{k}\right) }\geq 0$
for all $h\in {\cal S}\left( {\Bbb R}^{k}\right) $. Then it follows that $L$
is essentially selfadjoint on ${\cal S}\left( {\Bbb R}^{k}\right) \subset
L^{2}\left( {\Bbb R}^{k}\right) $, and the spectrum of $\bar{L}$ is positive.

\end{document}